\documentclass[12pt,preprint]{aastex}

\begin{document}


\title{Application of Monte Carlo Algorithms
to the Bayesian Analysis of the Cosmic Microwave Background }


\author{J. Jewell, S.Levin,  C.H. Anderson }
\affil{Jet Propulsion Laboratory,
Mail-Stop 126-347,  4800 Oak Grove Drive, \\
Pasadena, CA 91109-8099}



\begin{abstract}
Power spectrum estimation and evaluation of associated errors
in the presence of incomplete 
sky coverage; non-homogeneous, correlated instrumental noise; and foreground
emission is a problem of central importance for the extraction of cosmological
information from the cosmic microwave background.  We develop a
 Monte Carlo approach for the maximum likelihood
estimatation of the power spectrum.  The method is based on
an identity for the Bayesian posterior as a marginalization
over unknowns, and maximization of the posterior involves
the computation of expectation values as a sample average from
maps of the cosmic microwave background and foregrounds
given some current estimate of the power spectrum or cosmological
model, and some assumed statistical characterization of the foregrounds.
Maps of the CMB and foregrounds are sampled by a linear transform
of a Gaussian white noise process, implemented numerically with conjugate
gradient descent.  For time series data with $N_{t}$ samples,
and $N$ pixels on the sphere, the method has a computational expense
$KO[N^{2} + N_{t} \log N_{t}]$, where $K$ is a prefactor determined
by the convergence rate of conjugate gradient descent.
Preconditioners for conjugate gradient descent are given
for scans close to great circle paths, and the method allows partial
sky coverage for these cases by numerically marginalizing over
the unobserved, or removed, region.

\end{abstract}




\section{Introduction}
Power spectrum estimation and evaluation of the associated
errors in the presence of incomplete 
sky coverage; non-homogeneous, correlated instrumental noise; and foreground
emission is a problem of central importance for the extraction of cosmological
information from the cosmic microwave background.
From a Bayesian point of view, power spectrum estimation involves
the maximization of the posterior probability density,
with error bars given by the set of cosmological parameters
or power spectrum whose integrated posterior density achieves some
specified level of confidence.
A Bayesian approach to CMB analysis for large data sets
involving a direct evaluation of the likelihood 
is intractable due to the $O(N^{3})$ expense associated
with computing the inverse of
non-sparse matrices, or their determinants ( \cite{Borrill},
\cite{Bond1} ).
The goal of this paper is the development of alternative numerical methods,
specifically Monte Carlo
techniques, for  the Bayesian analysis of the CMB, including the
complications of incomplete sky coverage, correlated noise and foregrounds.

Previous work has demonstrated that
for a certain class of scanning strategies,
the signal and inverse noise matrices are block diagonal.
The block diagonal properties of these matrices
give an exact $O(N^{2})$ Bayesian method, and therefore tractable
for data sets as large as will be returned from Planck.
The complications of this method are that it cannot easily
accommodate partial sky converge or precessing scan strategies.
The method of \cite{OhSpergel}
computes the maximum of the likelihood through a Newton
Raphson method.  The numerical innovations of this method
involve Monte Carlo simulations and the use of conjugate gradient
descent, giving an overall expense $O(N^{2})$.
The method was proposed and numerically
demonstrated in the context of uncorrelated noise, and a region of
sky coverage of azimuthal symmetry, where a good preconditioner
can be constructed.  However, the algorithm is in fact
more general, provided there is sufficient memory for storage of
the needed matrices, and that conjugate gradient descent converges
quickly enough (i.e. there is a good preconditioner).

As suggested in \cite{WandeltRingSet}, we can use the
ring set approach to supply preconditioners.  An outstanding problem
to be solved is a way of retaining the mathematical advantages
of a ring set scan (block diagonal inverse noise and signal matrices)
while accommodating partial sky coverage.
The approach formulated in this paper handles the problem
of partial sky coverage by embedding the data in an azimuthally
symmetric region of sky, and using a Monte Carlo Markov Chain to
numerically marginalize over the unobserved part.
For scans close to ring sets, we therefore inherit good preconditioners,
allowing an extension of both the ring set and conjugate gradient methods
to scan strategies as planned for Planck.

For observations $y = s + f + \eta$,
where $(s,f,\eta)$ are the CMB signal, foregrounds, and noise respectively,
our approach to power spectrum estimation
is motivated by the identity (derived in Appendix 
\ref{AppPosteriorIdentity} )
\begin{equation}
\label{PosteriorIdentity}
\frac{p(\Gamma | y)}{p(\Gamma_{0} | y)}
 = \int d(s,f) \ \left[ \frac{p(\Gamma|s)}
{p(\Gamma_{0}|s)} \right] p(s,f|\Gamma_{0},y)
\end{equation}
where $\Gamma$ is any parameterization of conclusions
(such as the power spectrum or cosmological parameters), and
$\Gamma_{0}$ is any fixed guess.
The Bayesian posterior ratio on the left is given as an
integral over the unknown quantities which are assumed
to generate the observed data.
Maximization of the posterior involves computing the gradient
of equation \ref{PosteriorIdentity} which will be shown
to depend on the expectation value of the power spectrum
with respect to the random field $p(s , f |  \Gamma_{0},y)$,
\begin{equation}
\label{expectation}
E[C_{l}(s)|\Gamma_{0},y] = \int d(s,f) \ C_{l}(s) \ 
p(s,f |\Gamma_{0},y)
\end{equation}
We maximize the posterior ratio in equation \ref{PosteriorIdentity}
by the expectation maximization algorithm ( \cite{Dempster} )
which proceeds by iteratively setting
$C_{l}(\Gamma_{n+1}) = E[C_{l}(s)|\Gamma_{n},y]$.
The algorithm converges to the posterior maximum for a uniform
prior, and gives an un-biased, consistent estimator (see Appendix 
\ref{AppEstimator} ).  In this paper, we focus on computation of
the expectation value of the power spectrum
$E[C_{l}(s)|\Gamma_{n},y]$
given the data and some guess $\Gamma_{n}$ under the assumption
of perfect foreground separation (although we comment
on how the approach can be generalized to include foregrounds later
in the paper, and leave its numerical demonstration for future
work).

We compute the expectation value of the power spectrum
$E[C_{l}(s)|\Gamma_{n},y]$ numerically with
a Monte Carlo approach, where we sample maps of the CMB
from the probability density $p(s | f, y, \Gamma_{0} )$.
Conditioning on some estimate of the foregrounds, the method
exploits the fact that $p(s | f, \Gamma_{0},y)$ is a
Gaussian random field, and therefore completely characterized
by the mean field map and covariance matrix of fluctuations about that map.
Maps are sampled from $p(s | f, \Gamma_{0},y)$ by first
computing the mean field map with
conjugate gradient descent, and then sampling fluctuations
about the mean field map from a zero mean Gaussian field with
covariance matrix $(N^{-1} + C^{-1} )^{-1}$ (where
$N^{-1}$ is the inverse noise matrix, and $C^{-1}$ is the inverse
covariance matrix for the CMB).  These fluctuation maps
are sampled by a linear transformation, numerically
computed with conjugate gradient descent, of a
spatial white noise Gaussian process thereby generating maps
with all the same
statistical properties as samples from $p(s | f, \Gamma_{0},y)$.

Each step of conjugate gradient descent involves a multiplication
by the matrix $I + CN^{-1}$,
which can be done very quickly by multiplication by $N^{-1}$
in the basis in which it is diagonal, followed by a transform
to the spherical harmonic basis where $C$ is diagonal.
For spatially uncorrelated noise and circularly symmetric
beams, we only need to transform from the pixel to the spherical
harmonic domain, with an expense $O(N^{3/2})$  \cite{OhSpergel}.
In order to accommodate spatially
correlated noise, we transform to the time domain, followed by a
transform to the spherical harmonics, giving an expense
$K[O(N^{3/2}) + N_{t} \log N_{t} ]$ where $N_{t}$ is the number
of time samples, and $K$ depends on the convergence
rate of conjugate gradient descent.
Including the full complications of asymmetric beams,
we would need to compute a convolution on the sphere.
Using the convolution method of ( \cite{Wandelt00}), the expense of our
method is $K[O(N^{2}) + N_{t} \log N_{t} ]$.

The computational feasibility of this method is limited by
finding a numerical implementation of conjugate gradient descent
which converges quickly so that the prefactor $K$ above is small.
The strategy here is to embed the data in a region
covered by an exact ring set scan, following the intuition
that good preconditioners can be constructed for scan strategies close to
ring sets \cite{WandeltRingSet}.  Embedding the
data in a region on the sky with no observations (or where they
have been removed) is accommodated by numerically
marginalizing over the missing observations.  Moreover,
the same techniques can be used to marginalize over the foregrounds, and
provide Monte Carlo estimates of the confidence intervals for
cosmological parameters.

The paper is organized as follows.  We first review complications
with a direct computation of the likelihood, and provide an overview
of our approach.  We then discuss a technique 
we call transformed white noise sampling, which
allows us to sample maps representing fluctuations about the
mean field map for some guess of the power spectrum.
We demonstrate the method with a flat sky $512 \times 512$
test case, including incomplete sky coverage, with uncorrelated,
non-homogeneous noise.
We close with a discussion of further complications encountered in real
CMB experiments, and how they can be accommodated in the
framework presented here.

\section{Power Spectrum Estimation}
\subsection{Likelihood}
We begin with a brief review of the likelihood and complications
with its computational evaluation.  The data returned from an
experiment is a vector in the time domain $d(t)$, which is related
to the CMB signal on the sky $s(n)$ through some linear mapping
and additive Gaussian noise,
\begin{equation}
d(t) = \left[ \int dn' \ \delta[n(t)  - n'] \int dn'' \ B(n',n'')
s(n'') \right] + \eta(t)
\end{equation}
where $B(n',n'')$ is the beam of the instrument, and
$\eta(t)$ is Gaussian noise, assumed to be stationary
with a noise correlation matrix
$N(t,t') = N(t-t')$.  Denoting the linear mapping from time domain
to the sky as
\begin{equation}
R \equiv  \int dn' \ \delta[n(t)  - n'] \int dn'' \ B(n',n'')
\end{equation}
we will simply write $d=Rs + \eta$.
The likelihood for the power spectrum $C_{l}$ given the data is
\begin{equation}
p(d|C_{l}) \propto \int ds \ e^{- (d - Rs) N^{-1}_{t} (d-Rs) } e^{-sC^{-1} s}
\end{equation}
Any linear transformation of the data vector will generate
a Gaussian form for the likelihood (as reviewed in Appendix 
\ref{AppLikelihoods} ).
For example, we can transform the
data to an estimate of the CMB map
${\hat s} = (R^{T} N^{-1} R)^{-1} R^{T} N^{-1} d$ (as discussed
in detail in \cite{Tegmark1997}, \cite{Stompor} ).
The covariance matrix of this map is
\begin{eqnarray}
E[{\hat s} \otimes {\hat s}] & = &  C +  (R^{T} N^{-1} R)^{-1} 
\end{eqnarray}
which shows that the map ${\hat s}$ can be thought of as
signal $C$ with additive Gaussian noise with covariance matrix
$(R^{T} N^{-1}_{t} R)^{-1}$.  The likelihood can can equivalently
be written in terms of the map ${\hat s}(d)$ according to
\begin{eqnarray}
\log p[{\hat s}(d) | C_{l} ] & = &
- {\hat s}(d) [(R^{T} N^{-1} R)^{-1}  + C ]^{-1} {\hat s}(d)
- \log \det [(R^{T} N^{-1} R)^{-1} + C  ]
\end{eqnarray}
We can also write the likelihood in the original time domain,
\begin{equation}
\log p(d|C_{l} ) = - d (N + R C R^{T} )^{-1} d 
- \log \det [N + R C R^{T} ]
\end{equation}
Directly evaluating the likelihood in either the
spatial or time domain results in an $O(N^{3})$ computation,
as it involves inversion, or computation of the determinant,
of $[ (R^{T} N^{-1} R)^{-1} + C ]$ or $[N + RCR^{T}]$ respectively.
The computational expense is due to the fact that we do not
know the eigenbasis for either of these matrices,
and computing this basis is generally an $O(N^{3})$ problem.

The method of \cite{OhSpergel} solves the likelihood in the spatial domain
by evaluating the determinant with a Monte Carlo algorithm.
The method involves conjugate gradient descent to solve a linear
problem, and as such involves matrix multiplication, which carries
an $O(N^{2})$ expense.  The method was proposed and numerically
demonstrated in the context of uncorrelated noise, and a region of
sky coverage of azimuthal symmetry.  However, the algorithm is in fact
more general, provided there is sufficient memory for storage of
the needed matrices, and that conjugate gradient descent converges
quickly enough (i.e. there is a good preconditioner).

For a certain class of observing strategies, we can exactly
compute the likelihood, and for perturbations about these cases,
we can use the approximate case as a preconditioner (as suggested
in \cite{WandeltRingSet} ).
The approach formulated in this paper provides a consistent
way to do this, and involves a Monte Carlo Markov Chain approach to
numerically marginalizing over the unobserved part of azimuthally
symmetric regions of the sky.

\subsection{Embedding the Data - Marginalization}
The method to be developed in this paper involves embedding the
data in a region for which the signal and noise
matrices have desirable properties.  The likelihood for the
data in the context of some model is given as an integral over
the part of the embedding region which was not observed.
This gives the identity for the Bayesian posterior
for the power spectrum or cosmological model (denoted by $\Gamma$),
given the time domain data $d(t)$ as the integral
\begin{equation}
\frac{p(\Gamma | d)}{p(\Gamma_{0} | d)}
 = \int d(s^{(1)},s^{(2)},f) \ \left[ \frac{p(\Gamma|s^{(1)},s^{(2)})}
{p(\Gamma_{0}|s^{(1)},s^{(2)})} \right] p(s^{(1)},s^{(2)},f|\Gamma_{0},d)
\end{equation}
where we have explicitly written the CMB maps $s=(s^{(1)},s^{(2)})$
in terms of the part of the sky where we have data $s^{(1)}$ and
the complementary region $s^{(2)}$.
For the case of full sky coverage and prior knowledge $q(\Gamma)$, the log
posterior ratio is given as
\begin{equation}
\log \frac{p(\Gamma|s^{(1)},s^{(2)} )}{p(\Gamma_{0}|s^{(1)},s^{(2)} )} =
\log \frac{q(\Gamma)}{q(\Gamma_{0} )}  
- \sum_{l} (l+1/2) \left[ C_{l}(s^{(1)},s^{(2)})
 \left( \frac{1}{C_{l}(\Gamma)} - 
\frac{1}{C_{l}(\Gamma_{0})} \right)
+ \log \frac{C_{l}(\Gamma )}{C_{l}(\Gamma_{0})} \right]
\end{equation}
where the power at a given multipole order is
\begin{equation}
C_{l}(s^{(1)},s^{(2)}) = \frac{1}{2l+1}
\sum_{-l \le m \le l} 
\| \langle lm | s^{(1)} + s^{(2)} \rangle \|^{2}
\end{equation}
For other regions with azimuthal symmetry (an annulus or polar
cap), the posterior ratio would involve a similar form in terms
of block diagonal matrices.
The simple form of the log posterior ratio for full sky
coverage is one of the motivations for treating the problem of
partial sky coverage as missing data, and marginalizing over it
(detailed and justified in Appendix \ref{AppPartialCoverage}).

\subsection{Maximization}
In order to estimate the power spectrum, we would like to find
$\Gamma$ which maximizes the posterior given the noisy data.
Differentiating the posterior with respect to the parameters
$C_{l}$ gives a gradient in the direction
\begin{equation}
\left. \frac{\partial \log p(\Gamma|d)}{\partial C_{l}} \right|_{\Gamma_{0}}
\propto
\sum_{l} (1+1/2) \left[ \frac{E[C_{l} | y, \Gamma_{0}]}{C_{l}^{2}(\Gamma_{0})}
- \frac{1}{C_{l}(\Gamma_{0})} \right]
\end{equation}
where $E[C_{l}(s)|\Gamma_{0},d]$ was given previously in
equation \ref{expectation}.
An improvement of our current estimate of the power spectrum $\Gamma_{0}$
can then given according to a Newton-Raphson iterative scheme
( \cite{BJK} and \cite{OhSpergel} ) , where our current
guess is updated using an approximation to the curvature
of the likelihood ${\bf F}^{-1}_{ll'}[d,\Gamma_{n}]$, according to
\begin{equation}
C_{l}(\Gamma_{n+1}) = C_{l}(\Gamma_{n}) -
\frac{1}{2} \sum_{l'} {\bf F}^{-1}_{ll'}[d,\Gamma_{n}]
\left. \frac{\partial \log p(\Gamma|d)}{\partial C_{l}} \right|_{\Gamma_{n}}
\end{equation}
In Appendix \ref{AppEstimator}, it is shown that the
curvature matrix is given in terms of the expectation value
\begin{equation}
E [C_{l}  C_{l'} | d, \Gamma_{0} ]  = \int ds \ 
 C_{l}(s) C_{l'}(s) \ p(s|d, \Gamma_{0})
\end{equation}
In practice, we might want to avoid computing the inverse of the
curvature matrix, and simply use the diagonal elements.

For this paper, we instead implemented the simpler (although more slowly
converging) expectation maximization algorithm \cite{Dempster}.
This method essentially follows from Jensen's inequality, giving the
lower bound to the posterior
\begin{equation}
\log \left( \frac{p(\Gamma | d)}{p(\Gamma_{0} | d)} \right) \ge
\log \frac{q(\Gamma)}{q(\Gamma_{0} )}  
- \sum_{l} (l+1/2) \left[ E[C_{l}(s)|\Gamma_{0},d]
 \left( \frac{1}{C_{l}(\Gamma)} - 
\frac{1}{C_{l}(\Gamma_{0})} \right)
+ \log \frac{C_{l}(\Gamma )}{C_{l}(\Gamma_{0})} \right]
\end{equation}
For a uniform prior, the lower bound is maximized by
$E[C_{l}(s)| \Gamma_{0}, y]$.
In Appendix B, we prove that this estimator iteratively
converges to the maximum of the posterior, and is a consistent
and unbiased estimator (for a uniform prior).

\subsection{Computing Expectation Values for the Power Spectrum}
In order to iteratively converge to the optimal, consistent
estimator of the power spectrum, we need to compute, for any current
guess of the power spectrum $\Gamma_{0}$, the expectation value
$E[C_{l}(s) | y, \Gamma_{0}]$ as defined in equation \ref{expectation}.
Defining the mean field map
\begin{equation}
{\hat s} = [N^{-1} + C^{-1}(\Gamma_{0})]^{-1} N^{-1} y
\end{equation}
and associated power spectrum estimate
\begin{equation}
C_{l}({\hat s}) = \frac{1}{2l+1} \sum_{m}
\| \langle {\hat s} | lm \rangle \|^{2}
\end{equation}
the expectation value can be written by integrating over
fluctuations about the mean field map $\xi = s-{\hat s}$ as
\begin{eqnarray}
E[C_{l}(s) | y, \Gamma_{n}]  & = & \frac{1}{2l+1} \sum_{m}
\int d \xi \ \langle {\hat s} + \xi | lm \rangle 
\langle lm | {\hat s} + \xi \rangle \
\frac{e^{- \xi [N^{-1} + C^{-1}(\Gamma_{n}) ] \xi  }}
{\int d\xi' \ e^{- \xi' [N^{-1} + C^{-1}(\Gamma_{n}) ] \xi'  } } \nonumber \\
\end{eqnarray}
Since $E[\langle lm | \xi \rangle ] = 0$, the expectation value is
\begin{eqnarray}
E[C_{l}(s) | y, \Gamma_{n}]  & = & C_{l}({\hat s})
+ \frac{1}{2l+1} \sum_{m}
\int d \xi \ \langle  \xi | lm \rangle 
\langle lm | \xi \rangle \
\frac{e^{- \xi [ N^{-1} + C^{-1}(\Gamma_{n}) ] \xi  }}
{\int d\xi' \ e^{- \xi' [ N^{-1} + C^{-1}(\Gamma_{n}) ] \xi'  }  }
\end{eqnarray}
We refer to these two terms as the mean field map power spectrum
estimate (known to be biased) and the correction term.
Analytically, we know that the correction term is given by
$(2l+1)^{-1}  \sum_{m} \langle lm | [ N^{-1} + C^{-1}(\Gamma_{n}) ]^{-1}
| lm \rangle$ however, it is intractable to compute and store
the matrix $[ N^{-1} + C^{-1}(\Gamma_{n}) ]^{-1}$.  Our strategy is
to compute the correction term with a Monte Carlo method
described below.

\subsection{Transformed White Noise Sampling}
Due to the computational intractability of computing
the matrix inverse $(N^{-1} + C^{-1})^{-1}$, our strategy is
to compute the expectation value of the correction term
from fluctuation maps $\xi$ sampled from the zero mean Gaussian
random field with covariance matrix $(N^{-1} + C^{-1})^{-1}$.
We could easily sample fluctuation maps $\xi$ if we could
compute the eigenvectors and eigenvalues of the
matrix $(N^{-1} + C^{-1})$, since in this basis the Gaussian
probability density for $\xi$ factors.
However, computing the eigenvectors and eigenvalues is
again an $O(N^{3})$ operation.

Because of these difficulties, we look for an alternative
way to sample maps.  Defining $\delta = (N^{-1} + C^{-1}) \xi$,
we can write the log density, up to the normalization constant, as
\begin{eqnarray}
- \xi (N^{-1} + C^{-1}) \xi & = &  - \delta (N^{-1} + C^{-1})^{-1} \delta
\end{eqnarray}
The transformed Gaussian process has the covariance matrix
$N^{-1} + C^{-1}$, making it easy to sample
from.  Specifically, we can sample maps from this Gaussian process
by drawing two independent white noise maps
$(\omega_{1},\omega_{2})$, and setting
$\delta = C^{-1/2} \omega_{1} + N^{-1/2} \omega_{2}$.
Since both white noise maps are drawn independently
from a zero mean Gaussian process, the resulting covariance matrix is
$E[\delta \otimes \delta] =  N^{-1} + C^{-1}$
(as discussed in Appendix \ref{AppSampling}).

The maps with the correct statistical properties
are $\xi = (N^{-1} + C^{-1} )^{-1} \delta$, which can be solved
numerically for a given map $\delta$.
A numerically stable implementation involves
setting $\xi = (I + C N^{-1})^{-1}C \delta$ (as also noted in
\cite{OhSpergel}), and using 
conjugate gradient descent to solve
\begin{eqnarray}
(I +  C N^{-1}  ) \xi & = & 
C^{+1/2} \omega_{1} + C N^{-1/2} \omega_{2}
\end{eqnarray}
The resulting maps $\xi$ have the correct statistical properties,
since $E[\xi \otimes \xi] = (N^{-1} + C^{-1} )^{-1}$ (see Appendix 
\ref{AppSampling} ),
allowing us to compute the correction term to the power spectrum
estimate of the mean field map as a sample average.

In order to actually sample fluctuation maps by transforming
a Gaussian white noise process, we need to obtain the Cholesky
decomposition of both the signal and noise matrices.
If we have observations with uncorrelated noise on the sky,
then $N^{-1/2}$ is known in the spatial domain.  However, the scan
strategy of the instrument will result in complicated correlations,
so that computing $N^{-1/2}$ is intractable.
The noise is simple in the time domain, which suggests that
instead of choosing white noise maps in the spatial domain,
we instead draw from white noise Gaussian processes in the
time domain, where we know $N^{-1/2}$ in the Fourier basis, followed by
a transformation to the sky, where we can operate with
the signal matrix $C$.

For a realization of a white noise process in the time
domain $\tau$ and a white noise map in the spatial domain $\omega$,
we can compute a fluctuation map according to
\begin{equation}
[I+C R^{T} N^{-1} R] \xi = 
 C^{1/2} \omega + C R^{T} N^{-1/2} \tau 
\end{equation}
where $N^{-1/2}$ is known in the Fourier basis associated with the
time domain.  In Appendix \ref{AppSampling}, this procedure is justified
with a proof that the covariance matrix of the fluctuation
maps is $E[\xi \otimes \xi] =
[C^{-1}+ (R^{T} N^{-1}_{t} R)^{-1}]^{-1}$.

\subsection{Computational Expense}
The overall computational expense is fixed, for each iteration
of the power spectrum estimate, by the expense of matrix
multiplication and number of iterations needed to converge with
conjugate gradient descent.  In order to multiply by the matrix
$C R^{T} N^{-1} R$ we need to:
\begin{itemize}
\item  Transform to the time domain with the matrix $R$.
\item  Compute a time domain FFT.
\item  Multiply by $N^{-1}$ - this is a diagonal matrix
in the time domain Fourier basis
\item  Compute a time domain inverse FFT.
\item  Transform back to the spatial domain with $R^{T}$.
\item  Compute a spherical harmonic transform.
\item  Multiply by $C$ - this is a diagonal matrix in the
spherical harmonic domain if the embedding region is the full sky.
\end{itemize}
For the case of
circularly symmetric beams, the convolution with the beam
is not needed when operating with the matrix $R$ or its
transpose, giving an expense
$K O[N^{3/2} + N_{t} \log N_{t}]$, where $N_{t}$ is the
number of time samples, and $K$ is the prefactor related to
the convergence rate of conjugate gradient descent.
For cases where the beam is not circularly symmetric, the convolution
with the beam would have to be computed, increasing the expense to
$K O[N^{2} + N_{t} \log N_{t}]$.

\section{Numerical Example}
The simulations presented here involve the assumption of
spatially uncorrelated, but non-homogeneous, noise, as shown by the
upper left in figure $1$.  We also restrict the
problem to power spectrum estimation from a small patch of sky, and
neglect curvature (and therefore work with discrete Fourier basis
instead of spherical harmonics).  Our goal with these numerical simulations
is to demonstrate the approach in action.  Future work
will involve numerical implementations on the sphere.

A CMB power spectrum was generated using CMBfast \cite{CMBfast},
followed by the creation of a full sky map on the sphere using
the {\bf synfast} routine in the HEALPIX package \cite{Healpix}.
A smaller patch of the sky was then selected, and projected on
a rectangular grid.  This map $s(n)$ was taken to be the noise free
map, as shown in the upper right of figure 1.
We then generated a noise map $\eta(n)$
by selecting independently at each pixel a
Gaussian random number with variance scaled as shown according
to the upper left of figure 1.  Noise was added to the
noise free map, and data then removed in a rectangular hole
(as shown in figure 1).
This was taken to be a simulated data set
$y = s + \eta$ with partial coverage of
the rectangular patch of sky.

The inverse noise matrix was
given in terms of the variance  at the ith pixel $\sigma^{2}_{i}$ as
\begin{equation}
N^{-1}_{ij} = \begin{array}{cl}
\delta_{ij} \sigma_{i}^{-2} & \mbox{for the observed region of sky} \\
0 & \mbox{elsewhere }
\end{array}
\end{equation}
As an initial estimate of the power spectrum, we
computed the the power spectrum of the noisy, incomplete data
(as computed in the
two-dimensional Fourier basis since we neglected curvature)
and subtracted the power spectrum of a single simulated noise map
(on the region of sky where we have data).
We then iteratively adjusted the power spectrum with the expectation
maximization as above until convergence.

We found that preconditioning was in fact necessary to achieve a reasonable
convergence rate.  Using the initial estimate of the power spectrum
$C(\Gamma_{0})$, and computing the diagonal elements of the
inverse noise matrix in the Fourier domain $\langle k | N^{-1} | k \rangle$
with Monte Carlo noise maps, gave the preconditioner
\begin{equation}
M_{k k'}^{-1} \equiv \delta_{k k'} 
[ I + C_{kk}(\Gamma_{0}) N^{-1}_{kk} ]^{-1}
\end{equation}
Conjugate gradient descent was then used to solve the linear equations
for the mean field and fluctuation maps
\begin{eqnarray}
M^{-1} [I + CN^{-1}] {\hat s} & = & M^{-1} CN^{-1} y  \nonumber \\
M^{-1} [I + CN^{-1}] \xi & = & M^{-1} \delta
\end{eqnarray}
where $\delta = C^{+1/2} \omega_{1} + CN^{-1/2} \omega_{2}$
was computed from two independently chosen
spatial white noise maps $(\omega_{1},\omega_{2})$,
and $N^{-1/2}$ vanished in the unobserved part of the sky, and
elsewhere given by $\sigma_{i}^{-1}$.
The result of iterating the algorithm to convergence is shown in
figure 2.  Uncertainties in the power spectrum estimate were
computed by Monte Carlo, in which new CMB maps were generated,
noise added, and the algorithm run again.

\section{Additional Problems}
The methods presented above can be generalized to handle other
additional problems faced in the Bayesian analysis of the
CMB.  We do not provide numerical examples, but briefly
include comments on how to use transformed white noise sampling
to estimate error bars and include foregrounds.

\subsection{Confidence Intervals from a Markov Chain}
The ability to sample maps of the CMB given some estimate of the
power spectrum can be used to construct a Markov Chain Monte Carlo
algorithm which converges to the Bayesian posterior $p(\Gamma | d)$
itself.  Previously, Markov Chain Monte Carlo techniques have been
proposed for the extraction of marginal densities for cosmological
parameters from approximate Bayesian posterior densities for
the power spectrum ( \cite{Christenson} , \cite{Knox} ,
\cite{Lewis} , \cite{RMartin} )

We want to construct a transition matrix for the
{\it joint} density of CMB maps and the power spectrum.  This
can be done by following the Metropolis Hastings algorithm, where
we first assume detailed balance
\begin{equation}
p(\Gamma_{1},s_{1} | y) T(\Gamma_{2} , s_{2} | \Gamma_{1}, s_{1} ; y)
= T(\Gamma_{1}, s_{1} | \Gamma_{2} , s_{2} ; y)
p(\Gamma_{2} , s_{2} | y)
\end{equation}
From the condition of detailed balance we see that
\begin{equation}
p(\Gamma  | y)  = \int d(s, \Gamma_{2}, s_{2} ) \
T(\Gamma , s | \Gamma_{2} , s_{2} ; y)
p(\Gamma_{2} , s_{2} | y)
\end{equation}
which shows that the Bayesian
posterior is the marginalized equilibrium density, generated
by repeatedly taking steps generated with the transition matrix
$T(\Gamma_{2} , s_{2} | \Gamma_{1}, s_{1} ; y)$.
Given any approximation to the joint density
$p(\Gamma_{2} , s_{2} | y)$, repeated application of the transition
matrix will reach the equilibrium density.

We can construct the transition matrix as follows.
We first assume the proposal density is independent of the past
$\rho(\Gamma_{2} , s_{2} | \Gamma_{1}, s_{1} ; y) =
\rho(\Gamma_{2} , s_{2} |  y)$ and given by
\begin{equation}
\rho(\Gamma_{2} , s_{2} |  y) \propto
e^{- (d - Rs) N^{-1} (d-Rs) - s C^{-1}[\Gamma_{2}] s  }
{\tilde p}(\Gamma_{2} | y)
\end{equation}
where ${\tilde p}(\Gamma_{2} | y)$ is any approximation to the
Bayesian posterior itself.  We therefore have, as before, the
conditional density
\begin{equation}
\rho (s | \Gamma_{2}, y) =
\frac{e^{- (d - Rs) N^{-1} (d-Rs) - s C^{-1}[\Gamma_{2}] s } }
{\int ds' \ e^{- (d - Rs') N^{-1} (d-Rs') - s' C^{-1}[\Gamma_{2}] s' } }
\end{equation}
and the marginal density for the power spectra
given by our approximation ${\tilde p}(\Gamma_{2} | y)$.
With this choice of proposal matrix, the acceptance matrix
can be chosen to be
\begin{equation}
a[s_{2}, \Gamma_{2} | s_{1}, \Gamma_{1}] = \min \left[ 1,
\frac{p(\Gamma_{2}, s_{2} | y)}{p(\Gamma_{1}, s_{1} | y)}
\frac{\rho(\Gamma_{1},s_{1} |  y )}
{\rho(\Gamma_{2} , s_{2} |  y)} \right]
\end{equation}
What makes this computationally
tractable is that we know the ratios
\begin{equation}
\frac{p(\Gamma_{2}, s_{2} | y)}{p(\Gamma_{1}, s_{1} | y)}
= \frac{e^{- (d-Rs_{2})N^{-1} (d-Rs_{2}) - s_{2} C^{-1}[\Gamma_{2}] s_{2}
- \log \| C(\Gamma_{2}) \| } q[\Gamma_{2}] }
{e^{- (d-Rs_{1})N^{-1} (d-Rs_{1}) - s_{1} C^{-1}[\Gamma_{1}] s_{1}
- \log \| C(\Gamma_{1}) \| } q[\Gamma_{1}] }
\end{equation}
as well as
\begin{equation}
\frac{\rho(\Gamma_{1},s_{1} | y)}{\rho(\Gamma_{2}, s_{2} | y)} =
\frac{e^{- (d - Rs_{1}) N^{-1} (d-Rs_{1}) - s_{1} C^{-1}[\Gamma_{1}] s_{1}  }
{\tilde p}(\Gamma_{1} | y)}
{e^{- (d - Rs_{2}) N^{-1} (d-Rs_{2}) - s_{2} C^{-1}[\Gamma_{2}] s_{2}   }
{\tilde p}(\Gamma_{2} | y)}
\end{equation}
which simplifies to the explicitly computable acceptance probability
\begin{equation}
a[s_{2}, \Gamma_{2} | s_{1}, \Gamma_{1}] = \min \left[ 1,
\frac{e^{- \log \| C(\Gamma_{2}) \| } q[\Gamma_{2}] }
{e^{- \log \| C(\Gamma_{1}) \| } q[\Gamma_{1}] }
\frac{{\tilde p}(\Gamma_{1} | y)}
{{\tilde p}(\Gamma_{2} | y)} \right]
\end{equation}
In summary, a single step of the Markov chain involves:
\begin{itemize}
\item Choose a new guess for the power spectrum from
${\tilde p}(\Gamma_{2} | y)$.
\item  Sample a map from $\rho(s_{2} | \Gamma_{2} ; y)$ according to
\begin{itemize}
\item  Compute the mean field map ${\hat s}_{2} = [R^{T} N^{-1} R +
C^{-1}(\Gamma_{2}) ]^{-1} R^{T} N^{-1} y$.
\item  For two independently sampled white noise maps
in the spatial and time domains $(\omega, \tau)$,
compute $\xi$ as the solution to
$(I + C R^{T} N^{-1} R) \xi = C^{+1/2} \omega_{1} +
R^{T} N^{-1/2} \tau$.
\item  Set $s_{2} = {\hat s}_{2} + \xi$.
\end{itemize}
\item  Accept the transition $(\Gamma_{1}, s_{1}) \rightarrow
(\Gamma_{2}, s_{2})$ with probability
$a[\Gamma_{2}, s_{2} | \Gamma_{1}, s_{1} ; y]$.
\item  Continue
\end{itemize}
For circularly symmetric beams, each step of the Markov chain has expense
$K O[N^{3/2} + N_{t} \log N_{t}]$, giving a total
expense $N_{MC} A^{-1}_{MC}  K O[N^{3/2} + N_{t} \log N_{t}]$, where
$N_{MC}$ is the number of realizations of $(\Gamma, s)$,
and $A^{-1}_{MC}$ quantifies the ``efficiency'' of the Markov chain
(suggestively denoted by $A^{-1}$ since we intuitively expect it
to vary inversely with the average acceptance probability).

By the efficiency of the Markov chain, we mean the number of
proposed moves that must be made until one is accepted.
This efficiency depends entirely on how good the approximation
to the posterior ${\tilde p}(\Gamma_{2} | y)$ actually is.
We can always write the acceptance matrix as
\begin{equation}
a[s_{2}, \Gamma_{2} | s_{1}, \Gamma_{1}] = \min \left[ 1,
\frac{p(s_{2} | \Gamma_{2}, y)}{p(s_{1} | \Gamma_{1},  y)}
\frac{p(\Gamma_{2}| y)}{p(\Gamma_{1} | y)}
\frac{\rho(s_{1} | \Gamma_{1}  y )}
{\rho(s_{2} | \Gamma_{2}   y)}
\frac{{\tilde p}( \Gamma_{1} |  y )}
{{\tilde p}(\Gamma_{2} |  y)} \right]
\end{equation}
However, by construction, this is exactly
\begin{equation}
a[s_{2}, \Gamma_{2} | s_{1}, \Gamma_{1}] = \min \left[ 1,
\frac{p(\Gamma_{2}| y)}{p(\Gamma_{1} | y)}
\frac{{\tilde p}( \Gamma_{1} |  y )}
{{\tilde p}(\Gamma_{2} |  y)} \right]
\end{equation}
so that if our approximation was exact, the acceptance matrix is
unity for all proposals, and we reduce to an expense
$N_{MC} K O[N^{3/2} + N_{t} \log N_{t}]$.
Provided that we have reasonable approximations to the posterior,
we can converge to the exact posterior with this method, and the
use of transformed white noise sampling.

We also briefly comment on some practical issues with this approach.
In order to run the Markov chain, we need to compute the
mean field map {\it for every proposed power spectrum}.  While
this is feasible, we can speed up the Markov chain by sampling
a grid of power spectra from our approximation ${\tilde p}(\Gamma|y)$,
and pre-computing the mean field maps for each $\Gamma$ on our grid.
Then, for any proposed power spectrum, we can start with the initial
condition for conjugate gradient descent with the closest grid point.

\subsection{Inclusion of Foregrounds}
We conclude with a brief discussion of
the inclusion of foregrounds.  The inclusion of the foregrounds
involves sampling both CMB and foreground maps given some
estimate of the CMB power spectrum and some prior for the foregrounds.
For a Gaussian prior for the foregrounds, the same approach
to sampling can be followed.

Including foregrounds the data for the jth frequency channel
is given by
\begin{equation}
d_{j}(t) =  A_{j0} R_{j}  s + \sum_{p}  A_{jp} R_{j} f_{p} + \eta_{j}
\end{equation}
where $R_{j}$ is the mapping from the sky to the time domain
for the jth frequency channel, $A_{j0}$ is the response of the
CMB at the jth frequency, and we sum over the foreground
components $f_{p}$.  For a statistical characterization of the
foregrounds $\beta$ and a guess of the CMB power spectrum $\Gamma$,
we need to sample from the density given by (up to normalization)
\begin{equation}
- \log p(s,f | d, \Gamma_{0} ,\beta) \sim 
(d - A_{0} Rs - ARf) N^{-1} (d- A_{0}Rs - ARf)
- s C^{-1}(\Gamma_{0}) s - \log p(f | \beta) 
\end{equation}
We need to compute the expectation value
\begin{equation}
E[C_{l} | d,\Gamma_{0} , \beta] = \int d(s,f) \ C_{l}(s) \
p(s,f | d,\Gamma_{0}, \beta)
\end{equation}
This expectation value can be numerically computed
by sampling maps $(s,f)$ by the time average of a Markov chain
with equilibrium density $p(s,f | d,\Gamma_{0}, \beta)$.
One legitimate way to construct such a Markov chain is to alternately
sample maps from the conditional densities
$p(s | f,d,\Gamma_{0}, \beta)$ and $p(f | s ,d,\Gamma_{0}, \beta)$.
We briefly comment on sampling maps from each of these conditional densities.

Given some estimate of the foregrounds, we need to sample maps
from the conditional density
\begin{equation}
- \log p(s|f,  d, \Gamma_{0} ,\beta) \sim 
(d - A_{0} Rs - ARf) N^{-1} (d- A_{0}Rs - ARf)
- s C^{-1}(\Gamma_{0}) s
\end{equation}
Sampling from $p(s | f,d,\Gamma_{0}, \beta)$ proceeds in essentially the same
way as described above, but generalized for multi-frequency
data.  Specifically, the mean field map includes the subtracted
response from the foreground estimate
\begin{equation}
{\hat s} = [C^{-1} + A_{0} R^{T} N^{-1} R A_{0}  ]^{-1}
A_{0} R^{T}  N^{-1} (d-Af)
\end{equation}
and the fluctuation maps include a time domain white noise sample
for each frequency channel
\begin{equation}
(I + C R^{T} A_{0} N^{-1} A_{0} R) \xi =
C^{1/2} \omega_{1} + \sum_{j} R^{T}_{j} A_{0j} N^{-1}_{j} \tau_{j}
\end{equation}
Note that the covariance matrix of the fluctuations is
\begin{equation}
E[\xi \otimes \xi] = (C^{-1} + R^{T} A_{0} N^{-1} A_{0} R)^{-1}
\end{equation}
where the proof depends on the independence of the
time domain white noise maps for every frequency channel.

Given some estimate of the CMB, we need to sample from the
conditional density
\begin{equation}
- \log p(f | s,  d, \Gamma_{0} ,\beta) \sim 
(d - A_{0} Rs - ARf) N^{-1} (d- A_{0}Rs - ARf) - \log p(f | \beta) 
\end{equation}
If the prior for the foregrounds $p(f | \beta)$ is Gaussian
\begin{equation}
- \log p(f|\beta) \sim f B^{-1}(\beta) f
\end{equation}
then we can also use transformed white noise sampling.
First, we compute the mean field foreground map,
\begin{equation}
{\hat f} = [B^{-1} + R^{T} A N^{-1} A R]^{-1}
R^{T} A N^{-1} (d-A_{0} s)
\end{equation}
and then sampling fluctuations $\xi_{p}$ according to the transform
of white noise samples
\begin{equation}
(I + B R^{T} A_{p} N^{-1} A_{p} R) \xi_{p} =
B^{1/2} \omega_{1} + \sum_{j} R^{T}_{j} A_{pj} N^{-1}_{j} \tau_{j}
\end{equation}
Multiplying by the foreground signal matrix $B$ is either
done in the pixel or spherical harmonic basis depending on
the basis in which it is sparse.

If we use a non-Gaussian prior
(such as the maximum entropy method, or other priors), then
we will need to employ more general sampling techniques to sample
foreground maps, such as
the Metropolis algorithm.  In this case the needed expectation
value is to be computed as the {\it time
average} from a Markov chain with equilibrium density
$p(s,f | y,\Gamma_{n})$.

\section{Conclusions}
The fundamental hurdle to
numerically implementing an exact Bayesian approach to
CMB analysis, including complications of
partial sky coverage, correlated noise, and foregrounds,
is finding efficient ways to solve the linear
problem $(I + C R^{T} N^{-1} R) \xi = \delta$ for any vector $\delta$.
Solving the linear equation has an expense
$K O[N^{3/2} + N_{t} \log N_{t}]$ for circularly symmetric beams,
and the algorithm provides a tractable approach provided
$K$ can be made small enough.  The strategy we presented in this
paper allows the data to be embedded in an azimuthally symmetric
region of the sky covered by a Wandelt ring set scan,
with the intuition that, provided the true scan of the
instrument is close enough to the exact scan, we inherit
good preconditioners.

We also commented on how the method of transformed white noise
sampling can be used in Monte Carlo Markov Chain for the
entire Bayesian posterior.  The feasibility of this approach
depends on a good approximation to the posterior itself.
Previous work has demonstrated several
computationally feasible, unbiased estimates of
the power spectrum and associated error covariance matrix.
Any of these methods could therefore be used, in principle,
to give an approximate posterior, so that a Markov chain
approach can be used as a final consistency check.

Future work will incorporate the foregrounds in the algorithm
presented here, generalized for multifrequency data.
Maximization of the likelihood of the power spectrum
given the data again leads to the computation of the expectation
value $E[C_{l} | d, \Gamma_{0}]$, but now the marginalization includes
the foregrounds as well.  If the prior for the foregrounds
is Gaussian, then we can also use transformed white noise sampling
to sample a new foreground map while conditioning on the CMB.
If the prior used is non-Gaussian, other sampling schemes can be used,
including Gibbs sampling or the Metropolis algorithm.

\acknowledgements
This research was carried out at the NASA Jet Propulsion Lab,
under NASA AISRP Grant, and support from the Long Wavelength
Center.  We also thank Eric Hivon and Ben Wandelt for interesting discussions
during the course of this work.

\appendix

\section{Identities for the Bayesian Posterior  }

\subsection{Identity for the Posterior \label{AppPosteriorIdentity}  }
The data returned from a CMB experiment is a vector $d(t)$ 
in the time domain, generated from scanning the CMB signal $s(n)$
and foregrounds $f(n)$
on the sky and adding independent Gaussian noise.
The Bayesian posterior is given directly as an integral over
unknown quantities by
\begin{equation}
p(\Gamma | d) = \int d(s,f) \ p(\Gamma,s,f|d)
\end{equation}
For the case of Gaussian random fields, this integral
can be done analytically (as will be discussed in
Appendix \ref{AppLikelihoods} ), but evaluation of the resulting
likelihood leads to computationally intractable matrix manipulations.
For any estimate $\Gamma_{0}$
\begin{equation}
p(\Gamma | d) = \int d(s,f) \ \left[ \frac{p(\Gamma,s,f|d) }
{p(\Gamma_{0},s,f|d)} \right] p(\Gamma_{0},s,f|d)
\end{equation}
Since we have
\begin{equation}
\frac{p(\Gamma,s,f|d) }
{p(\Gamma_{0},s,f|d)} = \frac{p(\Gamma|s,f,d) }
{p(\Gamma_{0}|s,f,d)}
\end{equation}
this becomes
\begin{equation}
p(\Gamma | d) = p(\Gamma_{0} | d) \int d(s,f) \ \left[ \frac{p(\Gamma|s,f,d)}
{p(\Gamma_{0}|s,f,d)} \right] p(s,f|\Gamma_{0},d) 
\end{equation}
where we have used $p(\Gamma_{0},s,f|d) = p(s,f|\Gamma_{0},d) p(\Gamma_{0}|d)$.
By the assumed independence of the noise,
the joint density over which we are marginalizing is
\begin{equation}
p(\Gamma,s,f|d) \propto p(d|s) p(s|\Gamma) 
p(f|\beta)  q(\Gamma)
\end{equation}
where $q(\Gamma)$ is a prior for the parameters, $p(f)$ is a prior
for the foregrounds, and
$p(d|s)$ is completely determined by the noise properties of the
instrument, the beam, and scan strategy.  Given some estimate of the noise
free CMB signal, the density for a new guess of the power spectrum
is independent of the data, $p(\Gamma|s,f,y) = p(\Gamma|s)$, as shown by
\begin{eqnarray}
p(\Gamma|s,f,d) & = & \frac{p(d|s) p(s|\Gamma) 
p(f)  q(\Gamma)}{\int d\Gamma' \ p(d|s) p(s|\Gamma') 
p(f)  q(\Gamma')} \nonumber \\
& = & \frac{ p(s|\Gamma) q(\Gamma) }
{\int d\Gamma' \ p(s|\Gamma') q(\Gamma') }
\end{eqnarray}
Therefore, for {\it any estimate} $\Gamma_{0}$, our identity now reads
\begin{equation}
\frac{p(\Gamma | d)}{p(\Gamma_{0} | d)}
 = \int d(s,f) \ \left[ \frac{p(\Gamma|s)}
{p(\Gamma_{0}|s)} \right] p(s,f|\Gamma_{0},d)
\end{equation}
or for the likelihood ratio
\begin{equation}
\frac{p(d | \Gamma )}{p(d | \Gamma_{0} )}
 = \int d(s,f) \ \left[ \frac{p(s | \Gamma )}
{p(s | \Gamma_{0} )} \right] p(s,f|\Gamma_{0},d)
\end{equation}

\subsection{Transformations of the Data and Associated Likelihood
\label{AppLikelihoods} }
The likelihood of the data, generally in the time domain for
some experiment, is
\begin{equation}
p(d|C) \propto \int ds \ e^{- (d-Rs)N^{-1}(d-Rs) } e^{-s C^{-1} s }
\end{equation}
We first consider the case of transforming the time ordered data
to a map according to the transformation
${\hat s} = (R^{T} N^{-1} R)^{-1} R^{T} N^{-1} d$.
The likelihood is the probability density evaluated at the data.
Therefore, we first need to find the probability density of
{\it maps} given by transforming the density for time ordered data,
according to
\begin{equation}
p[{\hat s} | C_{l} ] = \int \delta d \ 
\delta[{\hat s} - (R^{T} N^{-1} R)^{-1} R^{T} N^{-1} d] \
\int ds \ e^{- (d-Rs)N^{-1} (d-Rs) } e^{-s C^{-1} s}
\end{equation}
This is equivalently
\begin{eqnarray}
p[{\hat s} | C_{l} ] & \propto &  \int \delta d \ 
\delta[{\hat s} - (R^{T} N^{-1} R)^{-1} R^{T} N^{-1} d] \
\int ds \ e^{- (d-Rs)N^{-1} (d-Rs) } e^{-s C^{-1} s} \nonumber \\
& = &  \int \delta d \ \int dk \ e^{- i k \cdot {\hat s} }
e^{+i k \cdot (R^{T} N^{-1} R)^{-1} R^{T} N^{-1} d }
\int ds \ e^{- (d-Rs)N^{-1} (d-Rs) } e^{-s C^{-1} s} \nonumber \\
& = &  \int dk \  e^{- i k \cdot {\hat s} }
\int ds \ e^{-s C^{-1} s}
\int \delta d \ e^{+i k \cdot (R^{T} N^{-1} R)^{-1} R^{T} N^{-1} d }
e^{- (d-Rs)N^{-1} (d-Rs) } \nonumber \\
& = &  \int dk \   e^{- i k \cdot {\hat s}  }
 \int ds \  e^{+ i k \cdot s  } e^{-s C^{-1} s}
\int \delta d \ e^{+i k \cdot (R^{T} N^{-1} R)^{-1} R^{T} N^{-1} (d - Rs )}
e^{- (d-Rs)N^{-1} (d-Rs) } \nonumber \\
& = &  \int dk \   e^{- i k \cdot {\hat s}  } e^{- k (R^{T} N^{-1} R)^{-1} k }
 \int ds \  e^{+ i k \cdot s  } e^{-s C^{-1} s} \nonumber \\
& = &  \int dk \   e^{- i k \cdot {\hat s}  } e^{- k (R^{T} N^{-1} R )^{-1} k }
 e^{- k C k} \nonumber \\
& = &  e^{- {\hat s} [(R^{T} N^{-1} R )^{-1} + C  ]^{-1} {\hat s} }
\end{eqnarray}
Notice that there are many time series vectors which correspond
to the same spatial map, since we can add any vector such that
$R^{T} N^{-1} d' = 0$.  This is possible since $R^{T}d$ vanishing means
that averages at the same point vanish.

We can also write the likelihood in terms of the data in the
original time domain.
This can be done by writing the characteristic function
\begin{eqnarray}
C(k) & = & \int \delta d \ e^{-i k \cdot d}
\int ds \ e^{- (d-Rs) N^{-1} (d-Rs)} e^{-sC^{-1} s} \nonumber \\
& = & \int ds \ e^{-sC^{-1} s} \int \delta d \ e^{-i k \cdot d} \
 e^{- (d-Rs) N^{-1} (d-Rs)}  \nonumber \\
& = & \int ds \ e^{- i k \cdot Rs}
e^{-sC^{-1} s} \int \delta d \ e^{-i k \cdot (d - Rs)} \
 e^{- (d-Rs) N^{-1} (d-Rs)}  \nonumber \\
& = & e^{- k N k } \int ds \ e^{- i k \cdot Rs}
e^{-s C^{-1} s}   \nonumber \\
& = & e^{- k (N + R C R^{T} ) k } 
\end{eqnarray}
which shows that
\begin{equation}
p(d|C_{l} ) \propto e^{- d (N + R C R^{T} )^{-1} d }
\end{equation}

\subsection{Partial Sky Coverage    \label{AppPartialCoverage}  }
The likelihood for the CMB given the theory for partial sky coverage
can be written as the marginalization over the unobserved part of
the sky.  Denoting the CMB $s= (s^{(1)},s^{(2)})$ as the CMB in the
observed and unobserved regions of sky respectively, we have
\begin{equation}
p(s^{(1)}|\Gamma) = \int ds^{(2)} \ p(s^{(1)},s^{(2)} |\Gamma)
\end{equation}
Therefore, the posterior for partial sky coverage can be written
\begin{eqnarray}
p(\Gamma|y) & \propto & q(\Gamma) \int d(s^{(1)},f) \ 
p(y|s^{(1)},f) p(s^{(1)}|\Gamma) p(f|\beta)
\nonumber \\
& = & q(\Gamma) \int d(s^{(1)},f) \ p(y|s^{(1)},f) 
\left[ \int ds^{(2)} \ p(s^{(1)},s^{(2)} |\Gamma) \right] p(f|\beta)
\nonumber \\
& = & q(\Gamma) \int d(s^{(1)},s^{(2)},f) \ p(y|s^{(1)},f) 
p(s^{(1)},s^{(2)} |\Gamma)  p(f|\beta)
\end{eqnarray}
This gives the identity, explicitly written for arbitrary sky coverage,
\begin{equation}
\frac{p(\Gamma|y) }{p(\Gamma_{0}|y)} = \int d(s^{(1)},s^{(2)},f) \
\left[ \frac{p(\Gamma| s^{(1)},s^{(2)} )}
{p(\Gamma_{0} | s^{(1)},s^{(2)} ) } \right]
p(s^{(1)},s^{(2)},f|\Gamma_{0},y)
\end{equation}
Because $s \equiv (s^{(1)},s^{(2)})$ is supported on the full
sky, the log posterior ratio is
\begin{equation}
\log \frac{p(\Gamma|s )}{p(\Gamma_{0}|s )} =
\log \frac{q(\Gamma)}{q(\Gamma_{0} )}  
- \sum_{l} (l+1/2) \left[ C_{l}(s)
 \left( \frac{1}{C_{l}(\Gamma)} - 
\frac{1}{C_{l}(\Gamma_{0})} \right)
+ \log \frac{C_{l}(\Gamma )}{C_{l}(\Gamma_{0})} \right]
\end{equation}
and the conditional density from which we are to sample
$(s^{(1)},s^{(2)})$ is
\begin{equation}
- \log p(s^{(1)},s^{(2)} |\Gamma_{0},y) \sim
(y -s^{(1)}) N^{-1} (y-s^{(1)}) + 
(s^{(1)},s^{(2)}) C^{-1}(\Gamma) (s^{(1)},s^{(2)})
\end{equation}
This is equivalent to setting the inverse
noise matrix to zero for the part of the sky where
there is no data.

\subsection{No Data and No Noise Limits}
Two limiting cases of our identity involve
``no data''  and ``no noise''.
In the no data limit, the inverse noise matrix vanishes
everywhere, so that $p(s|y,\Gamma_{0}) \rightarrow p(s|\Gamma_{0})$
and the posterior is given by the prior, as shown by
\begin{eqnarray}
\frac{p(\Gamma|y)}{p(\Gamma_{0}|y)} & = &  \int ds \ 
\left[ \frac{p(\Gamma|s )}{p(\Gamma_{0}|s )} \right] p(s  | \Gamma_{0})
\nonumber \\
& = &  \int ds \ 
\left[ \frac{p(s|\Gamma ) q(\Gamma)}{p(s|\Gamma_{0} ) q(\Gamma_{0})} \right]
p(s  | \Gamma_{0})
\nonumber \\
& = &  \frac{ q(\Gamma)}{ q(\Gamma_{0})} \int ds \ p(s|\Gamma )
\nonumber \\
& = &  \frac{ q(\Gamma)}{ q(\Gamma_{0})}
\end{eqnarray}
In the noise free limit, the conditional density
$p(s|y,\Gamma_{0}) \rightarrow \delta (s-s_{true})$, independent
of our choice of $\Gamma_{0}$, so that we converge to the
noise-free posterior ratio.

\section{Properties of the Power Spectrum Estimator \label{AppEstimator} }
\subsection{Posterior Maximum}
Recall the the algorithm used involves iterating
\begin{equation}
C_{l}(\Gamma_{n+1}) = E[C_{l} | \Gamma_{n}, y]
\end{equation}
and therefore the fixed point satisfies
$C_{l}(\Gamma) = E[C_{l} | \Gamma, y]$.
We can prove that this estimator maximizes the posterior,
or equivalently the log posterior.  A direct calculation shows that
\begin{equation}
\frac{\partial}{\partial C_{l}(\Gamma) } \frac{p(\Gamma|y)}{p(\Gamma_{0}|y)}
= \int ds \ \left[ \sum_{l} (l+1/2) C_{l}(s) \left(
\frac{1}{C_{l}^{2}(\Gamma)} - \frac{1}{C_{l}(\Gamma)} \right) \right]
\frac{p(\Gamma|s)}{p(\Gamma_{0}|s)} p(s|y,\Gamma_{0})
\end{equation}
so that at the maximum
\begin{equation}
0 = \int ds \ \left[ \sum_{l} (l+1/2) C_{l}(s) \left(
\frac{1}{C_{l}^{2}(\Gamma)} - \frac{1}{C_{l}(\Gamma)} \right) \right]
 p(s|y,\Gamma_{0})
\end{equation}
At the maximum we therefore have
\begin{equation}
C_{l}(\Gamma_{0}) = \int ds \  C_{l}(s)  p(s|y,\Gamma_{0})
\end{equation}
which is identically the fixed point of the
expectation maximization algorithm.
Therefore, the iteration converges to the maximum of the
posterior (for a uniform prior).

\subsection{Estimating the Curvature of the Likelihood}
After we have computed the maximum likelihood estimator of the
power spectrum (or maximum posterior estimate), we want to
find a confidence interval.  
There are several ways this confidence interval
might be approximated.  One approach is to compute the inverse
curvature matrix of the likelihood, and take the diagonal
entries as an estimate of the error bars (refs).
We comment below on how this can be done with the same method
used to compute $E[C_{l} | y, \Gamma]$.

Approximating the likelihood as a Gaussian functional of the
$C_{l}(\Gamma)$ is equivalent to a second
order Taylor expansion of the log likelihood about the maximum.
As before we have the identity for the likelihood
\begin{eqnarray}
\frac{p(d|\Gamma)}{p(d|\Gamma_{0})} & = &
\int ds \ \frac{p(s|\Gamma)}{p(s|\Gamma_{0})} 
p(s|d,\Gamma_{0} )
\end{eqnarray}
It is more convenient to parameterize the likelihood ratio
in terms of $\theta_{l} = C_{l}^{-1}$, so that
when embedding the data on the full sky, we have
\begin{equation}
\log \frac{p(s|\Gamma)}{p(s|\Gamma_{0})} =
- \sum_{l} (l+1/2) \left( 
[ \theta_{l}(\Gamma) - \theta_{l}(\Gamma_{0})] C_{l}(s)
- \log \frac{\theta_{l}(\Gamma)}{\theta_{l}(\Gamma_{0}) } \right)
\end{equation}
Denoting the curvature matrix of the likelihood
\begin{equation}
{\bf F}_{ll'}[d,\Gamma_{0}] \equiv - \frac{\partial^{2} \log p(d|\Gamma)}
{\partial C_{l} \partial C_{l'} }
\end{equation}
we have the relation
\begin{eqnarray}
- {\bf F}_{ll'}[d,\Gamma_{0}] 
 & = & \frac{1}{C^{2}_{l}(\Gamma_{0})} \frac{\partial^{2} \log p(d|\Gamma)}
{\partial \theta_{l} \partial \theta_{l'} }
\frac{1}{C^{2}_{l'}(\Gamma_{0})}
\end{eqnarray}
The curvature of $\log p(d|\Gamma)$ evaluated at the maximum,
where $C_{l}(\Gamma_{0}) = E[C_{l} | d, \Gamma_{0}]$ is therefore
\begin{eqnarray}
\frac{\partial^{2} \log p(d|\Gamma)}
{\partial \theta_{l} \partial \theta_{l'} }
 & = &  - \delta_{ll'} (l+1/2) C^{2}_{l}(\Gamma_{0}) \nonumber \\
& & + (l+1/2)(l'+ 1/2) \left[
E [C_{l}  C_{l'} | d, \Gamma_{0} ] -
C_{l}(\Gamma_{0}) C_{l'}(\Gamma_{0}) \right]
\end{eqnarray}
where we have defined the expectation value
\begin{equation}
E [C_{l}  C_{l'} | d, \Gamma_{0} ]  = \int d\xi \ 
 C_{l}({\hat s} + \xi) C_{l'}({\hat s} + \xi) \
\frac{e^{- \xi [N^{-1} + C^{-1}(\Gamma_{0})] \xi}}
{\det |N^{-1} + C^{-1}(\Gamma_{0}) |}
\end{equation}
This can, in principle, be computed with the conjugate gradient descent method
of transforming samples from a white noise process.  Future
work will study the accuracy and convergence properties of estimating
the curvature matrix from transformed white noise sampling.

\subsection{Covariance Matrices \label{AppCovarianceMatrices} }
The correctness of the algorithm for power spectrum
estimation presented in this paper is established by proving
that the covariance matrix of two linearly transformed vectors
has some specific form.  For simplicity of notation,
we choose to denote the covariance matrix of two vectors
$(x,y)$ as the expectation value of the outer product of
the vectors $E[x_{i} y_{j}] \equiv E[x \otimes y]_{ij}$.
For two matrices $A$ and $B$, an identity used repeatedly
in computing covariance matrices is
\begin{equation}
E[(Ax) \otimes (By)] = A E[x \otimes y] B^{T}
\end{equation}
which is shown simply by checking for each matrix element
\begin{eqnarray}
E[(Ax) \otimes (By)]_{mn} & = & 
\sum_{ij} E[A_{mi} x_{i} B_{nj} y_{j}] \nonumber \\
& = & \sum_{ij} A_{mi} E[x_{i} y_{j} ] B^{T}_{jn}
\end{eqnarray}
One example of this identity is in computing the
expectation value of maps computed from time ordered data.
One form of making a map from time ordered data is given by
\begin{equation}
{\hat s} = (R^{T} N^{-1} R)^{-1} R^{T} N^{-1} d
\end{equation}
as mentioned in section 2 and in \cite{Tegmark1997}.
According to the identity above, the covariance matrix of this map is
\begin{eqnarray}
E[{\hat s} \otimes {\hat s} ] & = &
(R^{T} N^{-1} R)^{-1} R^{T} N^{-1} E[d \otimes d] 
(R^{T} N^{-1} )^{T} [(R^{T} N^{-1} R)^{-1}]^{T} 
\end{eqnarray}
Substituting $d=Rs + \eta$, this is
\begin{eqnarray}
E[{\hat s} \otimes {\hat s} ] & = &
(R^{T} N^{-1} R)^{-1} R^{T} N^{-1} E[(Rs + \eta)  \otimes (Rs + \eta)] 
N^{-1} R  (R^{T} N^{-1} R)^{-1} \nonumber \\
& = &
(R^{T} N^{-1} R)^{-1} R^{T} N^{-1} \left(  R E[s \otimes s] R^{T} 
+ R E[s \otimes \eta ] \right. \nonumber \\
& & \left. + E[\eta \otimes s] R^{T} +
E[\eta \otimes \eta ] \right)
N^{-1} R  (R^{T} N^{-1} R)^{-1} \nonumber \\
& = &
(R^{T} N^{-1} R)^{-1} R^{T} N^{-1} \left(  R C R^{T}  + N \right)
N^{-1} R  (R^{T} N^{-1} R)^{-1} \nonumber \\
& = & C + (R^{T} N^{-1} R)^{-1}
\end{eqnarray}
where we have used the independence of the signal and noise,
and also that both are zero mean processes.  This result is
also proven in Appendix B below using the characteristic function
of the transformed time ordered data.

\subsection{Consistency}
The estimator given by the expectation maximization algorithm
(also equivalent to the maximum likelihood estimator) is given by
\begin{equation}
E[s \otimes s | y] = \int ds \ [ s \otimes s ]  \
\frac{e^{- (y - Rs) N^{-1} (y -Rs) - sC^{-1} s} }
{\int ds' \ e^{- (y - Rs') N^{-1} (y -Rs') - s' C^{-1} s'} }
\end{equation}
For partial sky coverage, this is equivalent to embedding
the data on the full sky and marginalizing over it
\begin{equation}
E[s \otimes s | y] =  \int d(s, y^{(2)}) \ [ s \otimes s ]  \
\frac{e^{- (y^{(2)} - s) {\tilde N}^{-1} (y^{(2)} - s)}
e^{- (y - Rs) N^{-1} (y -Rs) - sC^{-1} s} }
{\int d(s',y') \ e^{- (y' - s') {\tilde N}^{-1} (y' - s')}
e^{- (y - Rs') N^{-1} (y -Rs') - s' C^{-1} s'} }
\end{equation}
where we have arbitrarily chosen the full sky inverse noise matrix
\begin{equation}
N^{-1} = \left[ \begin{array}{cc}
R^{T} N^{-1} R & 0 \\
0 & {\tilde N}^{-1} \end{array} \right]
\end{equation}
The expectation of the covariance matrix is then
\begin{equation}
E[E[s \otimes s | y]] =  \int dy \ \int d(s, y^{(2)}) \ [ s \otimes s ]  \
\frac{e^{- (y^{(2)} - s) {\tilde N}^{-1} (y^{(2)} - s)}
e^{- (y - Rs) N^{-1} (y -Rs) - sC^{-1} s} }
{\int d(s',y') \ e^{- (y' - s') {\tilde N}^{-1} (y' - s')}
e^{- (y - Rs') N^{-1} (y -Rs') - s' C^{-1} s'} }
\end{equation}
Consistency of the estimator is then shown by proving that
$C = E [E[s \otimes s | y]]$.

Using the augmented noise matrix (which now has an inverse
on the full sky), we can define the mean field map
${\hat s} = [N^{-1} + C^{-1}]^{-1} N^{-1} (y,y^{(2)})$, and
write
\begin{eqnarray}
E[E[s \otimes s | y]] & = &
(N^{-1} + C^{-1})^{-1} N^{-1} E[y \otimes y] N^{-1} (N^{-1} + C^{-1})^{-1}
\nonumber \\
& = &
C(N+C)^{-1}  (N+C)  (N+C)^{-1} C
\nonumber \\
& = &
C(N+C)^{-1}  C
\end{eqnarray}
The expectation value of the correction is $(N^{-1} +C^{-1})^{-1}$, so that
\begin{eqnarray}
E[E[s \otimes s | y]] & = & C
(N+C)^{-1}  C + (N^{-1} + C^{-1})^{-1}
\nonumber \\
& = &
C(N+C)^{-1}  C + N(N + C)^{-1} C
\nonumber \\
& = & C
\end{eqnarray}
Therefore, the expectation maximization algorithm converges
to the maximum likelihood estimator for a uniform prior, which
is also a consistent estimator for arbitrary sky coverage.

\section{Correctness of Transformed White Noise Sampling
\label{AppSampling}   }
As derived in section ? above, the algorithm to
sample maps from the Gaussian process with covariance
matrix $(N^{-1} + C^{-1})^{-1}$ is
\begin{itemize}
\item  Draw $(\omega_{1}, \omega_{2})$ from a white noise process.
\item  Compute $\delta = C^{+1/2} \omega_{1} + C N^{-1/2} \omega_{2}$.
\item  Compute $\xi = ( I + C N^{-1})^{-1} \delta$ with conjugate gradient
descent.
\end{itemize}
We can prove that for any $N^{-1}$, even one which
does not have an inverse (i.e. as in the case of partial coverage
of the chosen embedding region), we have
$E[\xi \otimes \xi] = (N^{-1} + C^{-1})^{-1}$.
From the above we have
\begin{equation}
\xi = (N^{-1} + C^{-1})^{-1} [ C^{-1/2} \omega' + N^{-1/2} \omega]
\end{equation}
which gives
\begin{eqnarray}
E[\xi \otimes \xi] & = & (N^{-1} + C^{-1})^{-1} E [ ( C^{-1/2} \omega' + 
 N^{-1/2} \omega)
\otimes ( C^{-1/2} \omega' +  N^{-1/2} \omega ) ]
(N^{-1} + C^{-1})^{-1} \nonumber \\
& = & (N^{-1} + C^{-1})^{-1}   \left(
C^{-1/2} E[\omega' \otimes \omega'] C^{-1/2} +
C^{-1/2} E[\omega' \otimes \omega] N^{-1/2}  \right. \nonumber \\
& & \left. + N^{-1/2} E[\omega \otimes \omega'] C^{-1/2}
+ N^{-1/2} E[\omega \otimes \omega ] N^{-1/2} 
\right) (N^{-1} + C^{-1})^{-1} \nonumber \\
& = & (N^{-1} + C^{-1})^{-1}  \left(
 C^{-1} + N^{-1}   \right)
 (N^{-1} + C^{-1})^{-1}   \nonumber \\
& = & (N^{-1} + C^{-1})^{-1} 
\end{eqnarray}
where by independence of the two white noise maps,
$E[\omega \otimes \omega'] = E[\omega] \otimes E[\omega']$
which vanishes since the white noise process is zero mean.
An important point to notice is that the matrices $N^{-1}$ or
$N^{-1/2}$ can be singular in the sense that they do not have inverses
on the full sky (i.e. are generated by incomplete scanning of the sky).
In fact $N^{-1/2} \omega$ vanishes in the null space of
$N^{-1}$ (where we do not have data).

For a realization of a white noise process in the time
domain $\tau$ and a white noise map in the spatial domain $\omega$,
we compute a fluctuation map according to
\begin{equation}
\xi = (I+C R^{T} N^{-1} R )^{-1} 
[ C^{1/2} \omega + C R^{T} N^{-1/2} \tau ]
\end{equation}
where $N^{-1/2}$ is known in the Fourier basis associated with the
time domain.  We can prove that the maps $\xi$ have the correct
covariance matrix, $E[\xi \otimes \xi] = (R^{T} N^{-1} R + C^{-1})^{-1}$
by the direct calculation
\begin{eqnarray}
E[\xi \otimes \xi] & = & (R^{T} N^{-1} R + C^{-1})^{-1} 
E [ ( C^{-1/2} \omega + R^{T} N^{-1/2} \tau )
\otimes ( C^{-1/2} \omega + R^{T} N^{-1/2} \tau ) ]
(R^{T} N^{-1} R + C^{-1})^{-1} \nonumber \\
& = & ( R^{T} N^{-1} R  + C^{-1})^{-1}   \left(
C^{-1/2} E[\omega \otimes \omega] C^{-1/2} +
+ R^{T} N^{-1/2} E[\tau \otimes \tau ] N^{-1/2} R
\right) (R^{T} N^{-1} + C^{-1})^{-1} \nonumber \\
& = & ( R^{T} N^{-1} R  + C^{-1})^{-1}   \left(
C^{-1} + R^{T} N^{-1}  R \right) (R^{T} N^{-1} + C^{-1})^{-1} \nonumber \\
& = & ( R^{T} N^{-1} R  + C^{-1})^{-1}   
\end{eqnarray}
where again, the cross terms vanish since
independence gives $E[\omega \otimes \tau] =
E[\omega] \otimes E[\tau]$, which vanishes since both are zero
mean processes.

\section{Embedding the Data in Azimuthally Symmetric Regions}
We can also find the likelihood for the data embedded in an
azimuthally symmetric region covered by a ring set.  As
discussed in \cite{WandeltRingSet}, we
represent the signal on the ring set with coefficients
$a_{m m'}$, so that
\begin{equation}
s(\theta,\phi) = \sum_{m m'} a_{mm'} e^{- i m \theta} e^{-i m' \phi}
\end{equation}
where both indices range from $- L_{max} \le m \le L_{max}$.
Therefore, the signal on the sky,
parameterized on the ring set, is given by a two-D inverse FFT.
For any specific power spectrum, the corresponding signal matrix on the
ring set covering the embedding region is block diagonal.
We denote the ring set covariance matrix $T = E[a \otimes a]$.

Given the noise free signal on the ring set $s(\theta,\phi)$,
any observed data
set is given by projecting into the sub-region where we have data,
\begin{equation}
d(t) = Qs + \eta(t)
\end{equation}
where $Q$ vanishes on the regions of the ring set where we have no
observations.
As before we would first compute the mean field and fluctuation
maps.  In order to sample fluctuation maps, we would again compute,
\begin{eqnarray}
(I +   T   Q^{T} N^{-1} Q ) \xi & = & 
T^{+1/2} \omega + {\cal C}  Q^{T} N^{-1/2} \tau
\end{eqnarray}
where $\tau$ is a time domain white noise process,
$\omega$ is a spatial white noise process on the full sky,
$T^{+1/2}$ is the Cholesky decomposition of the ring set covariance
matrix.

Although it is possible to compute the Cholesky decomposition of the
ring set covariance matrix, we might instead sample $C^{1/2} \omega$
on the full sky, and then project into the region covered by the ring set,
giving maps $R C^{1/2} \omega$.  Then the covariance matrix for these
maps is, using the usual identity
\begin{eqnarray}
E[( R C^{1/2} \omega) \otimes ( R C^{1/2} \omega)] & = &
R C^{1/2} E[\omega \otimes \omega]  C^{1/2} R^{T} \nonumber \\
& = & R C R^{T}
\end{eqnarray}
which is the correct covariance matrix on the ring set region.
The point is that we do not have to compute the Cholesky
decomposition of the signal matrix on the ring set, but can instead
transform white noise maps with the Choelsky decomposition
of the full sky signal covariance matrix
(diagonal in the spherical harmonic basis), and then project down
to the ring set region.

We can construct a preconditioner as follows.
Define the projection from the time domain to the
ring set following a Wandelt scan strategy as
$W^{T} N^{-1} W$.  The for the same time domain noise matrix
$N^{-1}$, we can use a preconditioner given by
$W^{T} N^{-1} W$ and define
\begin{equation}
M^{-1} = [ I + T W^{T} N^{-1} W  ]^{-1}
\end{equation}
It is shown in \cite{WandeltRingSet} that $W^{T} N^{-1} W$
is block diagonal on the ring set, so that $M^{-1}$ can be
computed in $O(N^{2})$ operations.  We can then solve
the linear equations for the mean field and fluctuation maps
\begin{eqnarray}
M^{-1} [ I + T Q^{T} N^{-1} Q ] {\hat s } & = & 
M^{-1}T Q^{T} N^{-1} d \nonumber \\
M^{-1} [ I + T Q^{T} N^{-1} Q ] \xi & = &
M^{-1} \left( T^{+1/2} \omega + T Q^{T} N^{-1/2} \tau \right)
\end{eqnarray}
For scans which are close to the ring set scan the intuition is that
conjugate gradient descent will converge quickly for the
above equations.

%
%

\newpage

\section{Figures}
\begin{itemize}
\item {\it Figure 1} - The top left image shows the variance of the noise
in each pixel.  The black region is an unobserved ``hole''.
The upper right plot is a noise free realization, and
the lower left is a mean field map given the simulated noisy
data.  The bottom
right shows a typical fluctuation map computed with
conjugate gradient descent.
\item {\it Figure 2} - Plot showing the power spectrum estimation after
$20$ iterations starting from a flat initial guess.  The green dots
represent the results of separate runs on each of 10 simulated data sets,
with each shade of green representing a different run.  For each data set
we have produced a new noise-free map (drawn from the theoretical power
spectrum given by the solid black line), and added inhomogenous noise and a
hole as shown in Figure 1.  The initial guess used was the expected (flat)
noise spectrum, multiplied by a random number between 0 and 1.  At low
$l$ the SNR is high, and the
spread in the dots is caused by noise and sample variance.  At high $l$,
the SNR is low and the spread in the dots is caused by variation in the
initial guess, resulting in an upper bound as shown in the plot.  Also
shown are sample power spectra for a single simulated data set (light blue
line), and noise-free map (dark blue line).
\end{itemize}

\newpage

\begin{figure}
\plotone{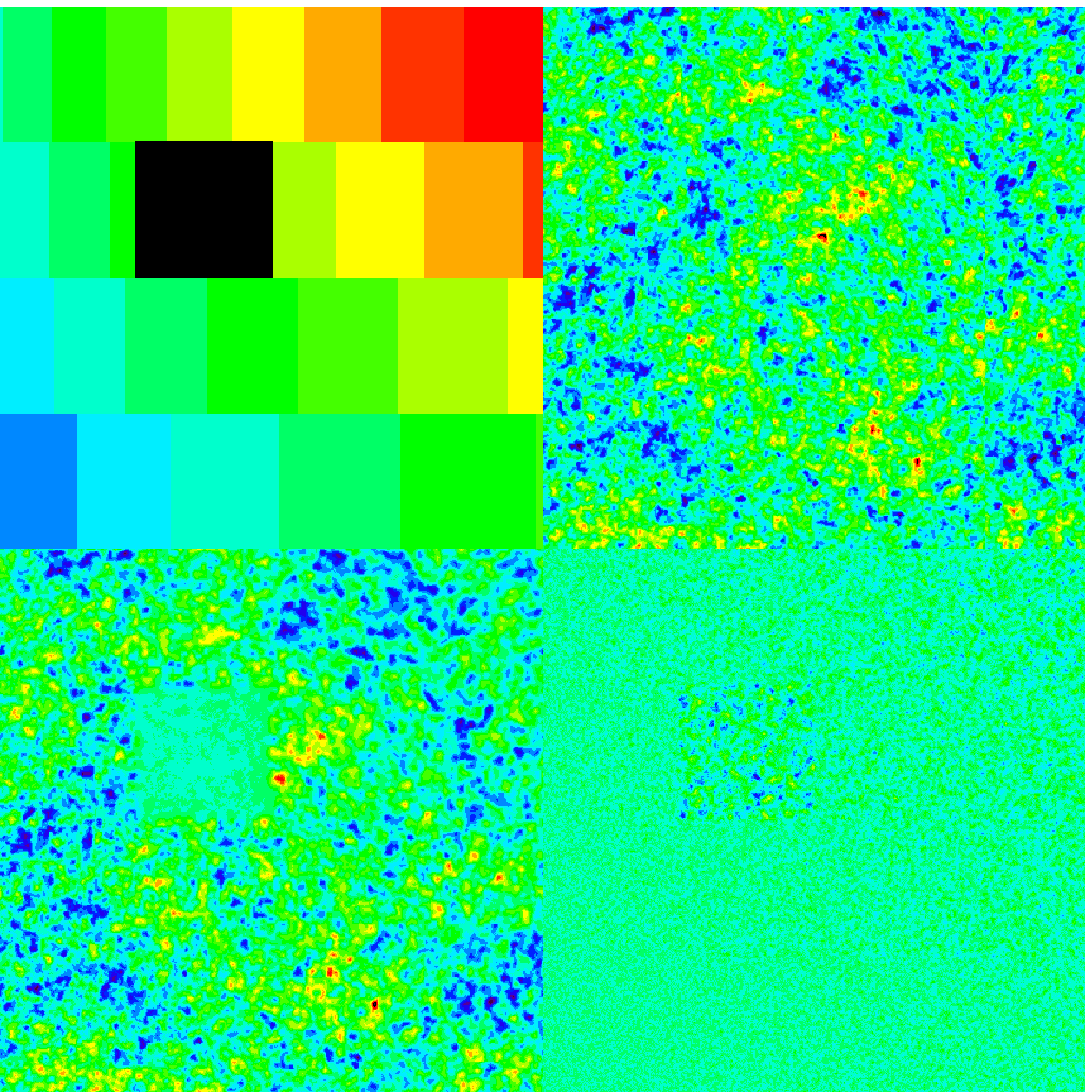}
\label{ExampleMaps}
\end{figure}

\begin{figure}
\plotone{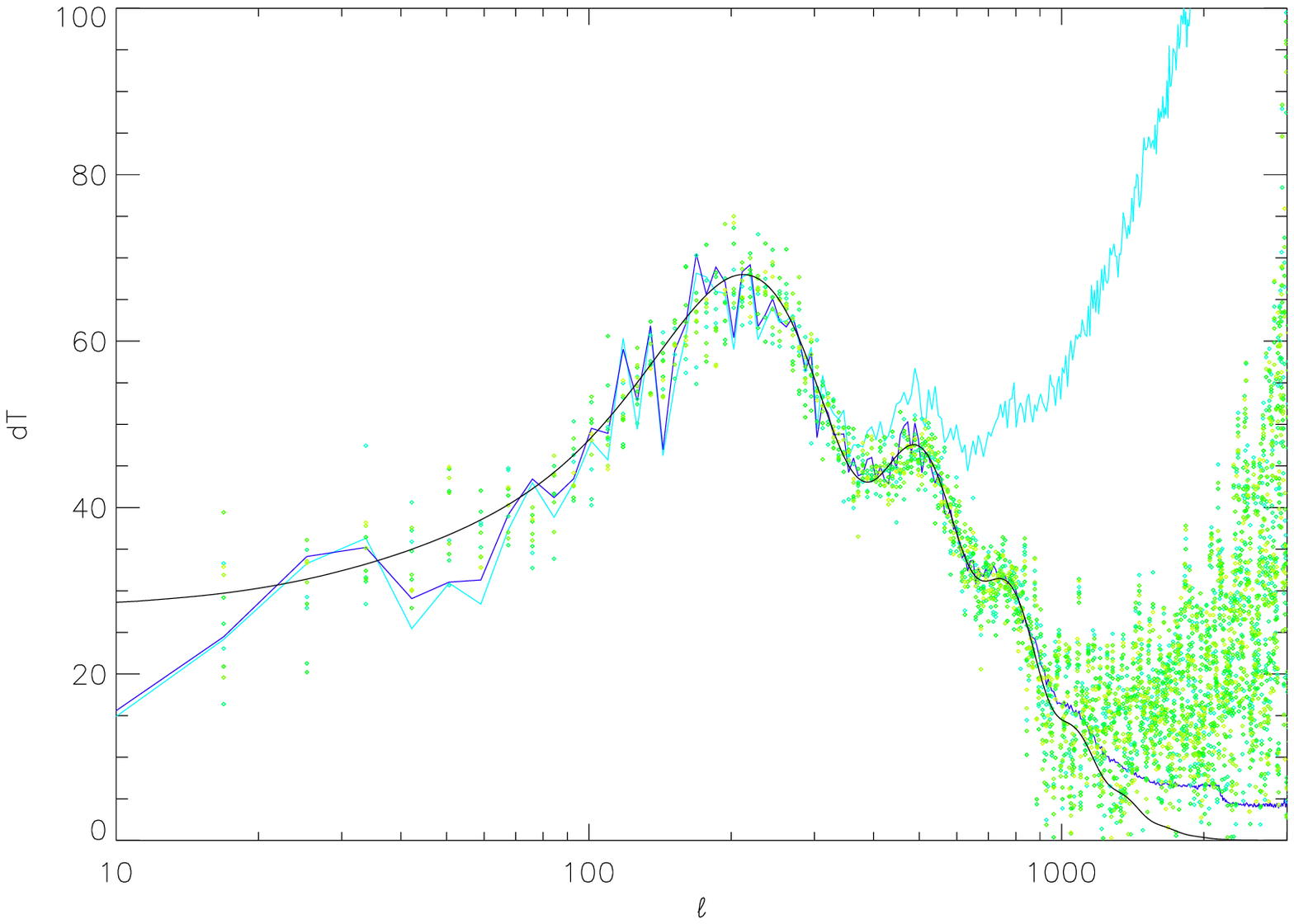}
\label{SpectrumEstimate}
\end{figure}

\end{document}